# Wheel of B2C E-commerce Development in Saudi Arabia

Rayed AlGhamdi (رائد الغامدي)[1], Anne Nguyen[2] and Vicki Jones[2]

[1]Faculty of Computing & IT, King Abdulaziz University
Jeddah, Kingdom of Saudi Arabia
raalghamdi8@kau.edu.sa

[2]School of ICT, Griffith University
170 Kessels Rd, Nathan, QLD 4111, Australia
a.nguyen@griffith.edu.au, v.jones@griffith.edu.au

**Abstract.** Online retailing (a model of B2C e-commerce) is growing worldwide, with companies in many countries showing increased sales and productivity as a result. It has great potential within the global economy. This paper looks at the current status of online retailing in Saudi Arabia, with particular focus on what inhibits or enables both the customers and retailers. It also analyses the status of Government involvement and proposes a layered model, known as the "Wheel of Online Retailing" which illustrates how Government intervention can benefit the e-commerce in Saudi Arabia.

**Keywords:** online retail, e-commerce, development, Saudi Arabia

## 1 Introduction

Although, the commercial industry is generally considered to be responsible for developing e-commerce within a country, the government should play a significant role in accommodating the e-commerce network particularly in: creating a favourable policy environment for e-commerce; and becoming a leading-edge user of e-commerce and its applications in its operations, and a provider to citizens of e-government services, to encourage its mass use [1].

Government support can take many forms from country to country. However, Government regulation can be critical in supporting e-commerce [2]. Comparing developing countries to developed countries, it is apparent that government incentive is most important in the developing world [2]. The aggressive driver for governments in some countries to take active role supporting e-commerce development is achieving economic success (e.g. Singapore, Taiwan, and Germany). In a rich developing nation like Saudi Arabia this is not such an aggressive motivator so government support is already taking its own unique form (and pace) but can still be guided effectively by the study of these successful cases.

Although Saudi Arabia has a large ICT marketplace, the growth of e-commerce activities is relatively slow [3,4]. The Saudi Government introduced e-commerce in 2001 in response to the fast expansion of e-commerce throughout the world. A permanent technical committee for e-commerce was established by the Saudi Ministry of Commerce. However, this Committee no longer exists, and from 2006, e-commerce supervision and development has been managed by the Ministry of Communications and Information Technology (MCIT). Unfortunately, there has been little progress since then.

## 2   E-government in Saudi Arabia

An e-government plan was set up with the following vision "By the end of 2010, everyone in the Kingdom will be able to enjoy from anywhere and at any time – world class government services offered in a seamless user friendly and secure way by utilizing a variety of electronic means" [5]. However, this vision has not been achieved as set up in a timely manner, which means the plan was not realistic [6]. The main problem, which was not taken into consideration, was the ICT infrastructure and assessing the e-readiness of the different government departments [7-11]. As a result, an e-government second action plan with the vision: "Enable use of efficient, integrated customer friendly and secure multiple e-Government services" (covering the period 2012-2016) has been launched; considering human resource training and development, promote cooperation and innovation culture, and maximizing efficiency of e-services provided by government agencies, [12].

E-government and e-commerce share some similarity in terms of transaction requirements. Therefore, development in e-government can serve as an engine to power e-commerce development [13]. The similarity between e-government and e-commerce is that both of them depend on ICT infrastructure, online payment systems and mailing/post systems to reach their users/customers and deliver their services/products [14].

## 3   Government role in e-commerce promotion

Government support takes various forms from country to country; however, government regulation can be critical to supporting e-commerce [2]. Online shopping shows rapid growth in the developed world. Significantly, the South Korean government has played a key role promoting e-commerce. The Malaysian government is encouraging small and medium enterprises (SMEs) to adopt e-commerce solutions, and in Australia, the Government is providing support is various forms [15].

## 4   Inhibitors and Enablers for the Customers and Retailers

The factors (inhibitors and enablers), which were detailed and discussed in [16-19], can be categorized into seven groups for the purpose of formulating possible explanation. The factors are categorized into the following groups: Cultural Issues (CI), Legislative Infrastructure (LI), Financial Infrastructure (FI), Logistics/Post Infrastructure (PI), ICT Infrastructure (II), Government Support (GS), and Internal Factors (IFs). From discussing the results, it is apparent that there is a distinct lack of government support of online retailing in Saudi Arabia

### 4.1 Cultural Issues (CI)

Online retailing and e-commerce in general is fairly new in the Saudi environment, and retailers are not selecting to sell online because their customers are not buying online [16,17]. Retailers in Saudi Arabia do not want to invest money opening online stores unless customers demand it. In contrast, a significant number of the customers surveyed say that, the fact that they are not familiar with the online retailers in Saudi Arabia is an inhibitor [18,19]. It looks like we are in vicious circle; customers say that there are not enough retailers online and retailers say there are not enough customers encouraging us to invest in online stores. For this reason, an intermediate third party is needed to facilitate change.

### 4.2 Legislative Infrastructure (LI)

The empirical evidence of this research continues to raise this problem as a major issue challenging online commerce in Saudi Arabia. Most of the transactions done by online businesses are paperless and signatures are digital. Consumers Protection - Currently, there is a consumer protection body that regulates traditional commerce. However, Saudis are frustrated that this body is not acting as expected and does not cover online commerce [20].

In Saudi Arabia, there is not yet any specific legislation related to e-commerce [21]. Both retailers and customers emphasized that the lack of clear regulations and legislation and the need for government supervision and support as being a significant inhibitor to their adoption of e-commerce.

### 4.3 Financial Infrastructure (FI)

In Saudi Arabia, many consumers are reluctant to use credit cards, both because of a lack of trust and because some consumers are culturally averse to carrying out transactions linked with conventional interest rates [22]. Therefore, it is apparent that provision of an alternative, trustworthy, and easy-to-use payment system has the potential to significantly influence the successful adoption of e-retailing in the KSA [21]. Some local banks in KSA, have started making normal ATM cards that are to be used as debit cards. Other possible solutions include the use of trusted third party payment systems, such as Paypal, which act as link between the credit card holders and the sellers.

Another option is the electronic bill presentment and payment system called SADAD that Saudi Arabia developed for billers and payers who are residents of the country. In essence, SADAD facilitates data exchange between registered billers and the nation's commercial banks. It relies on existing banking channels (such as Internet banking, telephone banking, ATM transactions and even counter transactions) to allow bill payers to view and pay their bills via their banks (for more details, see [22]). Currently, SADAD is limited to 100 billers only. This means that only the biggest billers in Saudi Arabia have access to this system.

### 4.4 Logistics/Post Infrastructure (PI)

Effective logistics infrastructure is a key for e-commerce. In Saudi Arabia the lack of housing mailboxes is an inhibitor to online retailing in Saudi Arabia. A program to accelerate allocation of house addresses and citizen adoption of mail boxes will have a positive effect on the adoption of e-commerce by providing the necessary infrastructure to support secure goods delivery to homes.

Unlike most developed nations, until 2005 nobody in Saudi Arabia had a house address to which deliveries could be made. Individuals had to subscribe to have a mailbox in the town's central post offices. In 2005, the new project for addressing and delivering mails to homes and buildings was announced and approved by the Saudi Post [3,23].

### 4.5 ICT Infrastructure (II)

The widespread availability of the Internet, especially broadband services, is an important indicator for strong ICT infrastructure [24]. Internet users in Saudi Arabia increased from one million (5% of the population) in 2001 to an estimated 13 million (46%) at the end of the third quarter of 2011 [25], while mobile broadband subscriptions reached 11.5 million, representing a penetration of 40.5% of the population, and the fixed broadband penetration rate stood at 30.6% of households for the same period [25]. However, most of the services provided by these companies were predominately from the main cities. Most small towns and villages are still not well served by Internet connections or have no Internet connections at all [20]. Internet connection fees in Saudi Arabia are considered high compared to the leading developing countries and developed countries [3]. Prices need to be reviewed to make it attractive for most of the households to connect to the Internet [20].

### 4.6 Government Intervention (GI)

The empirical evidence of this research suggests that the government interventions and support for e-commerce is very important factor in terms of flourishing e-commerce in Saudi Arabia [26]. The Saudi government does not give enough attention to e-commerce environment development [27].

While it is generally agreed that the commercial industry has to take the main responsibilities in developing e-commerce within a country, the government plays a significant role in promoting e-commerce particularly in (1) creating a favorable policy environment for e-commerce; and (2) becoming a leading-edge user of e-commerce and its applications in its operations, and a provider to citizens of e-government services, to encourage its mass use [1,28].

The empirical evidence of this research suggests that the government interventions and support for e-commerce is very important factor in terms of flourishing e-commerce in Saudi Arabia. It is ranked the second highest enabler for both customers and retailers.

The Saudi government does not give enough attention to e-commerce environment development. E-commerce requires an appropriate environment of legislative, financial, communicational, technological, and delivery infrastructures and building awareness among other issues [20].

The Saudi government has put its efforts towards e-government development [29]. E-government working plans and projects have been implemented to reach the vision "Enable use of efficient, integrated customer friendly and secure multiple e-Government services" by 2016 [20,21]. Surely, e-government and e-commerce share some similarity and well development in e-government can service as an engine to power e-commerce development as well [13]. However, e-commerce has different features which need to be considered

### 4.7 Internal Factors (IFs)

All the previously discussed issues: cultural issues, legislative infrastructure, financial infrastructure, logistics infrastructure, ICT infrastructure, and government support, seem external to the primary stakeholders, which require third parties: e.g. change agents, as Rogers [30] suggests, to take action. The factors discussed in this section are internal or related to individuals and organizations that make decisions to adopt change internally. The internal factors include: the lack of experiences for customers to purchase online and retailers to deal with e-commerce, customers looking for competitive prices, retailers' products are not suitable to be sold online, retailers' resistance to change, retailers distrust of e-commerce, e-commerce setup costs, perceived lack of profitability, and the difficulty for retailers to compete online.

## 5 A way forward: towards online retailing diffusion in KSA

The way toward online retailing diffusion in Saudi Arabia requires developing a strategic plan. The strategic planning for online retailing diffusion in Saudi Arabia is based on the results of this research.

- Very few retailers adopt e-commerce in Saudi Arabia
- Habit of Saudis in terms of buying online is the most influencing factor inhibiting retailers in Saudi Arabia to adopt e-commerce
- Lack of e-commerce experiences play significant role for both retailers and consumers to sell and purchase online
- There is not yet any specific legislation and regulations related to e-commerce in Saudi Arabia
- There is no government agency/body is clearly responsible for e-commerce in Saudi Arabia
- There is a need to develop trustworthy and secure online payment systems

A model called "*wheel of online retailing development in KSA*" is suggested in order to address the issues raised in this research and contribute to the development of online retailing, and e-commerce diffusion in Saudi Arabia.

The enhancement and development of these four areas (e-commerce legal framework, ICT infrastructure, logistic/ post infrastructure, and financial infrastructure) would enable the online environment and making it attractive for online businesses.

As Figure 1 illustrates, Government Support (GS) is placed in the external layer; legal framework, financial infrastructure, post infrastructure and ICT infrastructure are placed in the middle layer; cultural issues are placed in the second middle layer; and internal factors are placed in the central layer. The IFs are placed in the central layer because they are involved with the participants (i.e. customers and retailers) which means an individual (person/ organization) has to take the decision to change internally. However, the progress of the IFs is limited to the progress/ development/ maturity/ change in the surrounding layers (Cultural Issues (CI), Legislative Infrastructure (LI), Financial Infrastructure (FI), Logistics/Post Infrastructure (PI), and ICT Infrastructure (II)). Similarly, the progress/ development/ maturity/ change in CI, LI, FI, PI, and II are limited to the progress of government (the external layer).

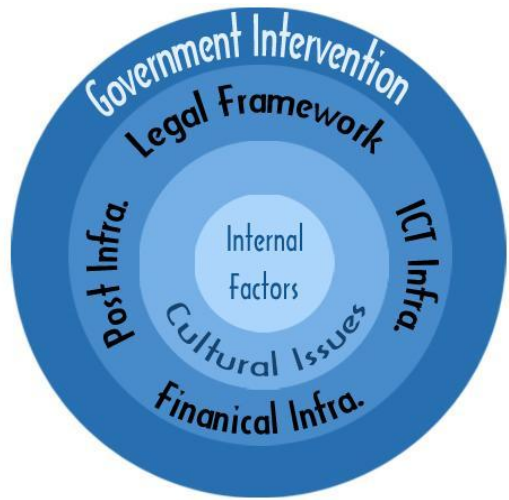

**Fig.1.** Wheel of online retailing development in KSA

In the light of the above findings, what actions could the government take to facilitate further development of online retailing in Saudi Arabia? There is little doubt that the clarification and enhancement of legislation and regulation in this area would be an appropriate priority for the government, as such matters are clearly part of its responsibilities. Similarly, further development of the IT infrastructure and domestic IT capabilities would benefit not only the e-retail industry but also the whole economy and the wider society. The government may also wish to facilitate or assist the development and growth of online payment systems (like SADAD) that would benefit consumers, businesses and the economy generally. A program to accelerate allocation of house addresses and citizen adoption of mail boxes and providing the necessary infrastructure to support secure goods delivery to homes is an urgent requirement not only for e-commerce but also to support all other e-services. The enhancement and development of these four areas (e-commerce legal framework, ICT infrastructure, logistic/ post infrastructure, and financial infrastructure) would enable the online environment and making it attractive for online businesses. There is no point to look at what inhibit consumers and businesses form purchasing or selling online while the environment is not ready to support this type of business. For this reason, a government body should be allocated for e-commerce development in Saudi Arabia. This government body then should work with the chambers of commerce and private sectors to sponsor solutions of e-commerce.

The second stage is to work on the cultural issues by meeting the concerns and contributes to resolve them. For example, lack of trust due to security and privacy would be change with the presence of customer protection law, old buying habit which dies hard and inhibit the growth of online shopping habit can be in somehow overcome by increasing the awareness and providing e-commerce educational programs. Even with the lack of products' physical inspection which is the most inhibiting factors for online shopping by consumers, leading retailers, good reputation vendors, brand name products would not encounter problems in this regard when they sell online. The empirical evidence shows that more than half of the consumers would buy online from leading/ good reputations/ brand names retailers. This indicates the opportunity of success for leading retailers, good reputation and strong brand name/familiar products firms entering online sales environment. Therefore, focusing on those retailers and encouraging them to adopt e-commerce would create an active online environment in Saudi Arabia for consumers making them more likely to be accepted to purchase online from them and motivate other retailer to enter online environment as this will create competitive pressure in the online environment.

## References

1. A. Scupola, "Government Intervention in SMEs' E-Commerce Adoption", M Khosrow-Pour (ed.), Encyclopaedia of Information Science and Technology, Second edn, IGI Global, pp. 1689-1695 (2009).
2. K. Kraemer, J. Dedrick, N. Melville and K. Zhu, "Global e-commerce: impacts of national environment and policy", Cambridge Univ Press, New York (2006).
3. S. Alfuraih, "E-commerce and E-commerce Fraud in Saudi Arabia: A Case Study", 2nd International Conference on Information Security and Assurance Busan, Korea, pp. 176-180 (2008).
4. CITC, "IT Report 2010 On the Internet Ecosystem in Saudi Arabia", Communications and Information Technology Commission, Riyadh (2010).
5. A. M. AL-Shehry, "Transformation towards E-government in The Kingdom of Saudi Arabia: Technological and Organisational Perspectives", The School of Computing, CCSR. De Montfort University (2008).
6. O. Alfarraj, S. Nielsen and L. Vlacic, "eGovernment initiatives and key factors causing the delay of their implementation in Saudi Arabia", 5th Conference on Qualitative Research in IT, Brisbane, Australia, pp. 130-141 (2010).
7. O. Alfarraj, S. Drew and R. AlGhamdi, "EGovernment Stage Model: Evaluating the Rate of Web Development Progress of Government Websites in Saudi Arabia", International Journal of Advances Computer Science and Applications, vol. 2, No. 9, pp. 82-90 (2011).

8. I. Abu Nadi, "Success Factors for eGovernment Adoption: Citizen Centric Approach", LAP LAMBERT Academic Publishing Gold Coast, Australia (2010).
9. I. Abu Nadi, L. Sanzogni, K. S. Sandhu and P. R.Woods, "Success Factors Contributing to eGovernment Adoption in Saudi Arabia: G2C approach", Saudi International Innovation Conference SiiC 2008 Proceeding, Leeds, UK, pp. 1-8 (2008).
10. M. Alshehri, S. Drew, and O. Alfarraj, "A Comprehensive Analysis of E-government services adoption in Saudi Arabia: Obstacles and Challenges". International Journal of Advanced Computer Science and Applications (IJACSA), vol 3, No. 2, pp. 1-6 (2012).
11. M. Alshehri, S. Drew, T. Alhussain, and R. Alghamdi, "The Effects of Website Quality on Adoption of E-Government Service: AnEmpirical Study Applying UTAUT Model Using SEM", in J Lamp (ed.), 23rd Australasian Conference On Information Systems (ACIS 2012), Melbourne, Australia, pp. 1-13 (2012).
12. Yasser eGov Program. "The e-Government Second Action Plan (2012 – 2016)". [cited 2012 7 March]; Available from: http://www.yesser.gov.sa/en/MechanismsandRegulations/strategy/Pages/-second_Implementation_plan.aspx. (2012).
13. C. Blakeley and J. Matsuura, "E-government: An engine to power e-commerce development", Proceedings of the European Conference on e- Government, Dublin, Ireland, pp. 39-48 (2001).
14. A. AL-Shehry, S. Rogerson, N. Fairweather and M. Prior, "The Motivations for Change Towards E-Government Adoption: Case Studies from Saudi Arabia", eGovernment Workshop, London, UK, vol. 6, pp 5-11 (2006).
15. DBCDE (Australian Deptment of Broadband Communication and the Digital Economy), "Background to the online retail forum", Department of Broadband, Communication and the Digital Economy, viewed 14 May 2011, <http://www.dbcde.gov.au/digital_economy/online_retail_forum/background_to_the_online_retail_forum> (2011).
16. R. AlGhamdi, S. Drew and W. Al-Ghaith, "Factors Influencing Retailers in Saudi Arabia to Adoption of Online Sales Systems: a qualitative analysis", Electronic Journal of Information System in Developing Countries (EJISDC), vol. 47, No. 7, pp. 1-23 (2011).
17. R. AlGhamdi, J. Nguyen, A. Nguyen and S. Drew, "Factors Influencing E-Commerce Adoption by Retailers in Saudi Arabia: A Quantitative Analysis", International Journal of Electronic Commerce Studies (IJECS), vol. 3, No. 1, pp. 85-100 (2012).
18. R. AlGhamdi, S. Drew and O. Alfarraj, "Issues Influencing Saudi Customers' Decisions to Purchase from Online Retailers in the KSA: A Qualitative Analysis", European Journal of Scientific Research, Vol. 55, No. 4, pp. 580-593 (2011).
19. R. AlGhamdi, A. Nguyen, J. Nguyen and S. Drew, "Factors Influencing Saudi Customers' Decisions to Purchase from Online Retailers in Saudi Arabia: A Quantitative Analysis", IADIS International Conference e-Commerce 2011, Roma, Italy, pp. 153-161 (2011).
20. F. Aleid, S. Rogerson and B. Fairweather, "A consumers' perspective on E-commerce: practical solutions to encourage consumers' adoption of e- commerce in developing countries - A Saudi Arabian empirical study", International Conference on Advanced Management Science, Chengdu, China, vol. 2, pp. 373-377 (2010).
21. S. Alwahaishi, A. Nehari-Talet and V. Snasel, "Electronic commerce growth in developing countries: Barriers and challenges", First International Conference on Networked Digital Technologies, Ostrava, Czech Republic, pp. 225 – 232 (2009).
22. SADAD, "About SADAD Payment System, viewed 10 Oct 2010, <http://www.sadad.com/English/SADAD+SERVICES/AboutSADAD/> (2004).

23. Saudi Post, "Saudi Post: Establishment and Development, Saudi Post", viewed 21 Nov 2009, <http://www.sp.com.sa/Arabic/SaudiPost/aboutus/Pages/establishmentanddevelopment.aspx>. (2008)
24. A. Sleem, "E-Commerce Infrastructure in Developing Countries", in S Kamel (ed.), Electronic Business in Developing Countries: Opportunities and Challenges, Idea Group Inc., USA, pp. 349-385 (2006).
25. MCIT (Saudi Ministry of Communication and Information Technology). "ICT indicators in K.S.A (H1-2011)". 2011 [cited 2011 11 Dec]; Available from: http://www.mcit.gov.sa/english/Development/SectorIndices/
26. R. AlGhamdi, S. Drew and M. Alshehri, "Strategic Government Initiatives to Promote Diffusion of Online Retailing in Saudi Arabia", in P Pichappan (ed.), Sixth International Conference on Digital Information Management, Melbourne, Australia, pp. 217-222 (2011).
27. R. AlGhamdi, S. Drew and A. Abugabah, "Designing Government Strategies to Facilitate Diffusion of Online Commerce: A focus on KSA", Journal of Information & System Management, vol. 1, No. 3, pp. 97-106 (2011).
28. S. Padmannavar, "A Review on E-Commerce Empowering Women's", International Journal of Computer Science and Telecommunications, vol. 2, No. 8, pp. 74-78 (2011).
29. S. Alkhalaf, S. Drew, R. Alghamdi, O. Alfarraj, "E-Learning System on Higher Education Institutions in KSA: Attitudes and Perceptions of Faculty Members". Procedia-Social and Behavioral Sciences, vol. 47: pp. 1199-1205 (2012).
30. E.M. Rogers, "Diffusion of Innovations", 5$^{th}$ ed., New York: Simon & Schuster, (2003).